# SECURE AND ROBUST IPv6 AUTOCONFIGURATION PROTOCOL FOR MOBILE ADHOC NETWORKS UNDER STRONG ADVERSARIAL MODEL

Zohra Slimane[1], Abdelhafid Abdelmalek[1],
Mohamed Feham[1] and Abdelmalik Taleb-Ahmed[2]

[1]STIC Laboratory University of Tlemcen Algeria
a_abdelmalek@mail.univ-tlemcen.dz

[2] LAMIH Laboratory University of Valenciennes and Hainaut Cambrésis France
Abdelmalik.taleb@mail.univ-valenciennes.fr

*ABSTRACT*

*Automatic IP address assignment in Mobile Ad hoc Networks (MANETs) enables nodes to obtain routable addresses without any infrastructure. Different protocols have been developed throughout the last years to achieve this service. However, research primarily focused on correctness, efficiency and scalability; much less attention has been given to the security issues. The lack of security in the design of such protocols opens the possibility of many real threats leading to serious attacks in potentially hostile environments. Recently, few schemes have been proposed to solve this problem, but none of them has brought satisfactory solutions. Auto-configuration security issues are still an open problem. In this paper, a robust and secure stateful IP address allocation protocol for standalone MANETs is specified and evaluated within NS2. Our solution is based on mutual authentication, and a fully distributed Auto-configuration and CA model, in conjunction with threshold cryptography. By deploying a new concept of joint IP address and public key certificate, we show that, instead of earlier approaches, our solution solves the problem of all possible attacks associated with dynamic IP address assignment in MANETs. The resulting protocol incurs low latency and control overhead.*

*Keywords*

*Mobile Ad hoc Networks, Auto-configuration, IP address, security, Threshold Cryptography, TCSAP, NS2*

## 1. INTRODUCTION

Spontaneous Mobile Ad hoc Networks (MANETs) are very attractive for either military or civilian applications in environments where fixed infrastructures are not available and rapid deployment is desired. In the last decade, large research efforts have been made to address challenges posed by MANETs. These challenges include mainly IP address auto-configuration, routing, security and QoS issues. In security context, the major part of research up to now was concentrated mainly on trust models and routing security problems. However, the lack of security in previously suggested auto-configuration schemes can lead to serious attacks in potentially hostile environments, mainly IP spoofing attack, sybil attack, traffic overload DoS attack, exhaustion address space attack, and conflict address attack. This problem was tackled by some few papers [1]-[5]. We have analyzed these proposals and pointed out their weaknesses and shortcomings in [13]; we have identified also the imperative security requirements related to this problem. In the present paper, we propose a new robust and secure stateful IP address allocation protocol for MANETs, by applying a cooperative security scheme to cope with malicious nodes including misbehaving nodes that could be compromised by potential adversaries. The scheme relies on a fully distributed Certification Authority based trust model in conjunction with a threshold signature scheme for issuing and revoking certificates, and 'On-line Joint IP Address and Public Key Certificate' model; this solves definitively the problem of some attacks such as IP spoofing and Sybil attacks, unsolved up to now by conventional mechanisms.

The remainder of the paper is organized as follows. In section 2, we present our trust model. Section 3 describes our threshold based auto-configuration scheme. Section 4 is devoted to the design of the basic building blocks of the protocol TCSAP which implements our proposal scheme. A security discussion is given in section 5. Section 6 presents our simulation experiments. Finally, section 7 concludes the paper.

## 2. TRUST MODEL

We adopt for our solution a Threshold Cryptography based Fully Distributed On-line Certification Authority (On-line CA) [7, 8]. In a (K,N) threshold MANET cryptosystem where K is the threshold and N the size of the MANET, each node member of the MANET holds a share of the CA's private key. The On-line CA service is achieved transparently by a large enough subset of nodes (K out of N). In a spontaneous MANETs, the threshold cryptosystem can be implemented as follows:

a) To generate randomly and in a distributed manner (without a trusted party) a pair of CA's private/public keys, to split the CA's private key among the network and to allow shareholders to verify the correctness of their shares. To do so, we use the joint verifiable random secret sharing protocol [9] based on Shamir's secret sharing [10].

b) To provide for any new joining node with a share of the Network's private key [7]

c) To provide a threshold digital signature scheme to sign issued, renewed or revoked certificates. With regard to the threshold signature protocol, a variety of discrete log based schemes have been proposed [11] including Nyberg-Rueppel or ElGamal- like and Elliptic Curve threshold digital signatures.

Note that in our scheme, each node member of the MANET must hold: On one hand, a valid share of the CA's private key and a pair of private/public keys approved by the On-line CA.

Nodes in the MANET need to be able to verify at any time whether a public key is revoked. They need also to be able to revoke either their own public keys or the public keys of malicious/compromised nodes. We assume a revocation scheme implemented within the MANET based on 'accusation' using threshold signatures [16]. Each node holds three tables *RCL* (Revoked Certificates List), *ANL* (Accused Nodes List), and *BL* (Black List).

We assume the trust model proposed here implemented in the module called DPKI (Distributed Public Key Infrastructure, see Figure 1).

## 3. AUTO-CONFIGURATION MODEL

Let us consider a standalone MANET. We develop a stateful auto-configuration scheme inspired from MANETconf [6] which we call TCSAP (Threshold Cryptography based Secure Auto-configuration Protocol) . We distribute the auto-configuration service to all nodes (size= N) in such a way that only a threshold number K of nodes can collaborate in performing the service functionality. Then, the IP Address for a newly arrived node is assigned by a subset of at least K nodes. Instead of MANETconf scheme in which an affirmative response from all nodes in the network is needed before assigning any available IP address to a newly arrived node, our scheme modifies MANETconf protocol and saves the communication bandwidth by assigning free IP addresses without asking permission from any other node in the MANET. To achieve this, we divide the Address Space into a fixed number (say M) of disjoint IP Address Blocks with equal sizes. We define an IP Address Block as a consecutive range of IP addresses. The parameter M is a power of 2.

### 3.1. State Information

At any instant of time, each configured node must maintain some state information defined hereafter:

• Free IP Address Table (FAT): contains the lowest free address of each IP Address Block. (i-e M values)

• Pending IP Address Table (PAT): contains recently assigned IP addresses which are not registered yet.

• Registered IP Address Table (RAT): Each entry in this table contains any assigned and registered IP address, the corresponding node's identity, its public key and the On-line joint certificate validity period. A registered node will be removed from the RAT if its certificate has expired. Nodes wishing to maintain their addresses must make a request for maintenance within a time specified before the expiry of their certificates.

• Requester Counter (RC): this counter is maintained for each new node requesting for an auto-configuration and to which an IP address is assigned but not registered yet. It is incremented for each new request. To prevent the Exhaustion Address Space Attack, the authorized attempts for the Auto-configuration Service Requesting are limited.

A configured node must update his state information in the following situations:

(i) Each time it reboots,

(ii) Each time it leaves and joins the MANET again,

(iii) If it has not been solicited for a long time to perform the Auto-configuration Service.

The node wishing to update its state information must collect redundant data from at least K nodes.

### 3.2. IP Address Allocation

A new node will be assigned randomly one of the lowest free addresses contained in the FAT, which means that the IP address is assigned in an increasing order from a randomly chosen IP Address Block. We impose to the new joining node to obtain its IP address from at least K nodes. We use for this purpose the threshold signature described above and the new concept of 'On-line Joint IP address and Public Key Certificate'. Each allocated IP address in the network is bound to node's identity by means of this certificate which must be signed by the On-line CA.

After having received a signed IP Address, the new node must broadcast a signed registration message to all nodes to be able to participate actively in the network. Any assigned IP address which is not registered yet is removed from the FAT and kept in the PAT, either by the K signer nodes after having assigned this address or by all nodes after having received a registration message for a higher IP address in the same IP Address Block. If the registration message is received the IP address is removed from the PAT and put in the RAT.

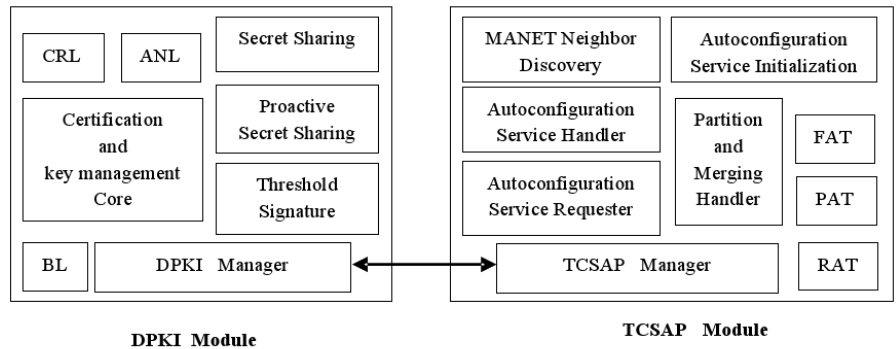

Figure 1. Functional Blocks of TCSAP and DPKI modules

## 4. TCSAP PROTOCOL DETAILS

This section specifies our new protocol TCSAP (Threshold Cryptography based Secure Auto-configuration Protocol) which implements the trust model and the auto-configuration scheme described above. We provide a 'building block approach' based decomposition of the protocol, each having its own specific functionality. The properties of these functional components and their interaction define the overall behavior of the protocol. Obviously, TCSAP module interacts also with DPKI module (Figure 1).

### 4.1. Definition of Node's States

- Unconfigured node: any node wishing to join the MANET, and which is not already registered with an 'On-line Joint IP address and Public Key Certificate'.

- Configured node: any registered node within the MANET with an 'On-line Joint IP address and Public Key Certificate'

- Node with Configuration in Progress: any unconfigured node which has initiated an auto-configuration process that is not finished yet.

### 4.2. Neighbors Discovery Component

The first component of our protocol is the Neighbors Discovery component. A node will discover it's one-hop neighbors by broadcasting periodically signed Discovery messages. The signature here is done according to the node's Off-line public key certificate. The node uses the DiscoveryTimer to detect the presence of its one-hop neighbors. This timer is rescheduled each time a Discovery_Request message is broadcast.

#### 4.2.1. Type of Messages

- *Discovery_Request:* this message is used by an *unconfigured* node in the *Neighbors Discovery* protocol to discover its one-hop neighbors.
- *Discovery_Reply:* is a reply message to any *Discovery_Request* message when the responder is in state *unconfigured.*
- *Discovery_Welcome:* is a reply message to any *Discovery_Request* message when the responder is in state *Configured.*

211

International Journal of Computer Networks & Communications (IJCNC) Vol.3, No.4, July 2011**4.2.2. Neighbors' Discovery Protocol**

The Neighbors Discovery protocol is executed automatically by a node on boots/reboots when his state is Unconfigured. The Discovery_Request message must contain the originator's Off-line public key certificate and its signature. Depending on the state of the recipient, the replay to this message will be:

```
1 :   bool ready = true;
2 :   ready ∧ nodestate = unconfigured →
3 :      broadcast Discovery_Request;
4 :      start DiscoveryTimer;
5 :      ready := false;
6 :   receive Discovery_Request →
7 :      if (SigCheck(Discovery_Request)) then
8 :         send Discovery_Replay;
9 :      else
10:         discard message;
11:      fi
12:   receive Discovery_Replay →
13:      if (SigCheck(Discovery_Replay)) then
14:         save neighbor's data in cache;
15:      else
16:         discard message;
17:      fi
18:   receive Discovery_Welcome →
19:      if (SigCheck(Discovery_Welcome)) then
20:         stop DiscoveryTimer;
21:         nodestate := configuration in progress;
22:         ready := true;
23:         start Autoconfiguration Service Requesting;
24:      else
25:         discard message;
26:      fi
27:   receive Init_Start →
28:      if (SigCheck(Init_Start)) then
29:         stop DiscoveryTimer;
30:         nodestate := configuration in progress;
31:         ready := true;
32:         process Autoconfiguration Service Initialization;
33:      else
34:         discard message;
35:      fi
36:   timeout (DiscoveryTimer) →
37:      if ( |F_N| ≥ k ) then
38:         nodestate := configuration in progress;
39:         start Autoconfiguration Service Initialization;
40:      else
41:          erase cache;
42:          nodestate := unconfigured;
43:          ready := true;
44:      fi
```

Figure 2. Algorithm 1: Neighbors Discovery Protocol

*a) Node state: Unconfigured*

The receiving node (say $P_i$) checks the signature, if it is valid it saves the originator's identity and its certificate in the one-hop Neighbors List $(N_L)_i$, and replies by a *Discovery_Reply* message containing its Off-line public key certificate, its $(N_L)_i$ and the signature, otherwise it discards the message.

212

International Journal of Computer Networks & Communications (IJCNC) Vol.3, No.4, July 2011

In this process, the requester stores in its cache, within a timeout period determined by the *DiscoveryTimer,* all its one-hop neighbors and their associated data, checks the neighbors lists intersection ($F_N$ : set of Founding Nodes):

$$F_N = \bigcap_{P_i \in H}(N_L)_i \qquad (1)$$

where $H$ is the set of all one-hop neighbors including the requester node itself.

If $|F_N| \geq K$ then the requester executes the *Auto-configuration Service Initialization* protocol, otherwise it clears its cache and retry another Neighbor's Discovery attempt.

### b) Node state: Configured

The recipient checks the signature; if it is valid it replies by a *Discovery_Welcome* message containing its IP address, its Off-line public key certificate and the signature, otherwise it discards the message. When the requester receives the *Discovery_Welcome* message, it concludes that a MANET is already established and has to start the *Auto-configuration Service.*

### c) Node state: Configuration in Progress

In this state, the node will discard all MANET Discovery messages.

The algorithm for the *Neighbors Discovery protocol* using the Abstract Protocol Notation [12] is given in (Figure 2).

## 4.3. MANET Initialization

The MANET is initialized by at least K neighbor nodes. When an *Unconfigured* node has discovered a number equal or greater than K neighbors belonging to the same neighborhood it starts the *Auto-configuration Service Initialization*. The founding neighbors have to jointly perform this initialization (Figure 3).

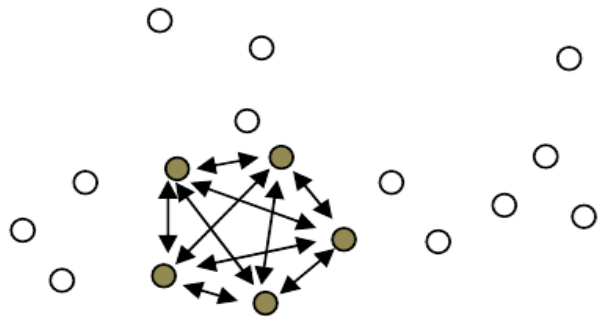

Figure 3.  Network Start up with 5 founding neighbors (*K* = 5)

### 4.3.1. Type of Messages

- *Init_Start:* this message is used to start the initialization of the MANET.
- *Init_Ack:* this message is broadcast by a founding node as replay and acknowledgment to *Init_Start* message.

- *Init_Advert:* is an advertisement message used to inform all nodes in the network that an *Auto-configuration service Initialization* is being processed.
- *Init_Oppos:* is an Opposition message to *Init_Start* and *Init_Advert* message, used when an Auto-configuration Service exists already in the MANET.

### 4.3.2. Initialization protocol

Our scheme is based on IPv6 addressing. Since we target standalone MANETs, we shall use site-local addresses (Figure 4).

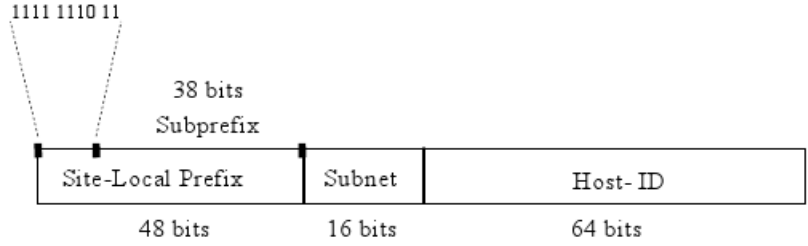

Figure 4. IPv6 site-local address

Host-ID with 64 bits allows a number of possible IP addresses greater than $18.10^{18}$, which is sufficient even for long-lived and large scale MANETs.

The site-local prefix is a binary sequence starting with FEC, FED, FEE or FEF (in hexadecimal notation), that is to say a prefix of FEC:: /48 up to FEF:: /48. Thus, the prefix begins always with the following ten bits "1111 1110 11"; we call sub-prefix the remaining 38 bits of the prefix.

The *Auto-configuration Service Initialization* is carried out in three phases (the algorithm is shown in Figure 5):

*a. Selection of the network prefix and attribution of IP addresses to founding nodes*

In this phase, first the founding neighbor nodes have to select the sub-prefix (38 bits) and the subnet (16 bits). Then, they have to set up their Host-ID (64 bits).

• *Selection of sub-prefix and subnet*

Let us set $N_{sp} = 2^{38}$ and $N_{sn} = 2^{16}$. The first node finding the threshold subset $F_N$ of the Founding neighbors, informs others nodes belonging to $F_N$ by broadcasting a signed *Init_Start* message including the list of these nodes.

Each $P_i \in F_N$ chooses randomly $x_i \in Z_{N_{sp}}$ and $y_i \in Z_{N_{sn}}$ (where $Z_{N_{sp}}$ and $Z_{N_{sn}}$ are cyclic groups under addition modulo respectively $N_{sp}$ and $N_{sn}$).

Then it broadcasts them in a signed *Init_Ack* message. The above broadcasts must be achieved in a limited time interval fixed by the *InitTimer*. After receiving all values $x_i$ and $y_i$, each $P_i \in F_N$ computes: $subprefix = \sum_{P_i \in F_N} x_i (\mod N_{sp})$ and $subnet = \sum_{P_i \in F_N} y_i (\mod N_{sn})$

International Journal of Computer Networks & Communications (IJCNC) Vol.3, No.4, July 2011

Two or more auto-configuration service initializations could to be processed simultaneously at different places in the network (Figure 6). To avoid this scenario, we provide the following solution: just after receiving the list of the Founding nodes, the node with lowest identity $U_i$ broadcasts a signed *Init_Advert* message to inform all nodes in the network. This message will be ignored except for the two following situations:

```
 1:  bool ready = true;
 2:  nodestate = unconfigured →
 3:     broadcast Init_Start;
 4:     if (my identity is the lowest among F_N) then
 5:        broadcast Init_Adv;
 6:        nodestate := configuration in progress;
 7:        broadcast Init_Ack;
 8:        start InitTimer;
 9:        ready := false;
10:     else
11:        nodestate := configuration in progress;
12:     fi
13:  receive Init_Oppos →
14:     if (SigCheck(Init_Oppos)) then
15:        nodestate = unconfigured;
16:        return;
17:     else
18:        discard message;
19:     fi
20:  receive Init_Adertv from node P_i∈ F_N ∧ ready →
21:     if (SigCheck(Init_Advert)) then
22:        if (P_i,s identity is the lowest) then
23:           broadcast Init_Ack;
24:           nodestate := configuration in progress;
25:           start InitTimer;
26:           ready := false;
27:        else
28:           continue;
29:        fi
30:     else
31:        discard message;
32:     fi
33:  receive Init_Advert from node P_i∉ F_N →
34:     if (SigCheck(Init_Advert)) then
35:        if (P_i,s identity < the lowest identity in F_N) then
36:           nodestate = unconfigured;
37:           return;
38:        else
39:           continue;
40:        fi
41:     else
42:        discard message;
43:     fi
44:  receive Init_Ack from node P_i∈ F_N →
45:     if (SigCheck(Init_Adv)) then
46:        save Init_Ack data from P_i in cache;
47:     else
48:        discard message;
49:     fi
50:  timeout (InitTimer) →
51:     if (each Init_Ack from P_i∈ F_N has been deceived) then
52:        compute subprefix and subnet;
53:        self-assign IP Address;
54:        call DPKI Manager;
55:        (secret sharing and certificates signing)
56:     else
57:        nodestate = unconfigured;
58:        return;
59:     fi
```

Figure 5. Algorithm 2: Initialization Protocol

International Journal of Computer Networks & Communications (IJCNC) Vol.3, No.4, July 2011

- If the recipient is in the state *Configured* it replies by an *Init_Oppos* message indicating that the network has been already initiated.
- If the recipient is participating in an auto-configuration service initialization, it will stop the process if the lowest identity among the local Founding nodes is greater than the identity of the advertising node; otherwise it continues the initialization process.

- *Host –ID setup*

Since prefix and subnet are common values for all IP Addresses, only the Host-ID space is divided into M blocks according to section 3. In addition to the list of the lowest free Host-ID of each block, the *FAT* will contain the corresponding subnet (required for network partition and merge management. M must be greater than the number of founding nodes ($M > |F_N|$).

Founding nodes ($P_i \in F_N$) are classified in an ascending order according to their identities $U_i$. Each $P_i \in F_N$ will assign itself the lowest value in the block corresponding to its order in the classification. Thus, the node with lowest $U_i$ gets the lowest value of block 1, the next gets the lowest value of block 2, and so on.

*b. Creation of the On-line Certification Authority*

When the above phase is finished, the *TCSAP Manager* calls the *DPKI Manager* to start the secret sharing protocol after which the On-line Certification Authority is created and its public key is signed.

*c. Signing the 'On-line Joint IP address and Public Key Certificate' for each founding node*

In this phase, each node $P_i \in F_N$ must get its 'On-line Joint IP address and Public Key Certificate' signed by at least *k* founding nodes. At this level, each node checks the correctness of the self-assigned Host-ID before performing a threshold signature. If this holds, then the MANET is completely initialized and all services can be started (Auto-configuration, routing, On-line Certification,…).

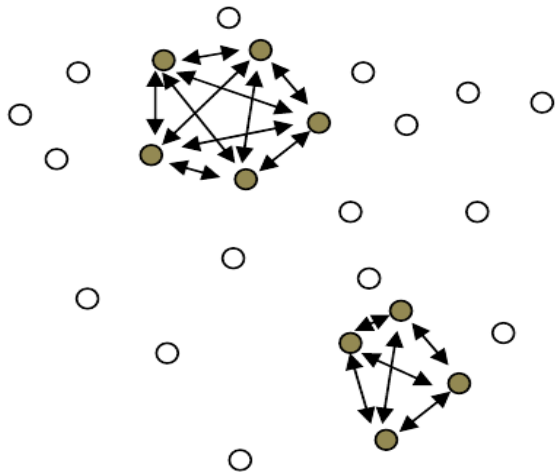

Figure 6. A scenario when two Initializations occur simultaneously ($|F_N| \geq K = 4$)

## 4.4. Auto-configuration of a newly arrived node

In the proposed scheme, a new joining node is assigned an IP address by means of the *Auto-configuration Service Requesting* (see component in Figure 1). Subsequently it is provided with an 'On-line Joint IP address and Public Key Certificate'. This is achieved in 4 phases: (1) research of co-signers, (2) selection of co-signers, (3) threshold signature of the On-line certificate and (4) MANET registration.

### 4.4.1. Type of Messages

- *Config_Request:* this message is used to start an *Autoconfiguration Service Requesting* after receiving a *Discovery_Welcome* message.
- *Config_Reply*: this message is used by a node in state *Configured* as a reply to *Config_Request* message.
- *Config_Cert_Request:* this message is used to request an 'On-line Joint IP address and Public Key Certificate'.
- *Config_Cert_Reply:* this message is used by the combiner as a reply message to *Config_Cert_Request* message.
- *Config_Advert:* this message is sent by the combiner to inform all nodes about the new assigned IP address.
- *Config_Alert:* this message is sent when a malicious node is discovered among the coalition.
- *Config_Register:* this message enables a new configured node to perform a registration within the MANET.
- *Config_Error:* this message is sent as an error reply to any *Config* type message

### 4.4.2. Research of Co-Signers

A new joining node which has received a valid *Discovery_Welcome* message as reply to its MANET *Discovery_Request* message starts automatically the *Autoconfiguration Service Requesting*. First, it assigns itself a link-local address formed by pre-pending the well known local prefix FE80::0/64 to the Host-ID of the IP address contained in any received *Discovery_Welcome* message. Since this Host-ID is unique, the obtained link-local address is also unique. Then, it broadcasts to all nodes in a predefined radius $r_k$ a *Config_Request* message including its 'Off-line Public Key Certificate' and its signature. It uses in the IPv6 packet his link-local address as source address, the site-local scope all-nodes multicasts address FF05::1 as the destination address, and a HopLimit of $r_k$ less than the threshold K. The algorithm for this procedure is given in Figure 7.

Any receiver of the message checks if the requester is not in the *BL*, then it checks the signature; if the requester is listed in the *BL*, or if the signature is not valid the message is discarded. Otherwise, it checks the *Requester Counter (RC)*. If the requester has already reached the limit of the authorized attempts for the *Autoconfiguration Service Requesting*, it is declared as malicious and the message is discarded. Otherwise, the recipient sends a *Config_Reply* message including the On-line Certification Authority's public key, the list of available subnets and the corresponding list of the lowest free Host-ID of each block from its *FAT*, the received HopLimit, its 'Off-line Public Key Certificate', and its signature.

The algorithm for the processing of the *Config_Request* message by a co-signer is shown in Figure 8.

```
1:  int r_k, HopLimit, numServers;
2:  numServers < k →
3:      self-assign link local IP Address;
4:      HopLimit := r_k;
5:      broadcast Config_Request;
6:      start ConfigTimer;
7:  receive Config_Replay →
8:      if (SigCheck(Config_Replay)) then
9:          save servers data in cache;
10:     else
11:         discard message;
12:     fi
13: timeout (ConfigTimer) →
14:     if (numServers < k) then
15:         if (r_k < k) then
16:             erase cache;
17:             r_k := r_k + 1;
18:         else
19:             nodestate = unconfigured;
20:             return;
21:         fi
22:     else
23:         start (coalition selection, certificate requesting and registration);
24:     fi
```

Figure 7. Algorithm 3: Research of Co-Signers

Note that any emitted or relayed IPv6 packet must be signed and the nodes have to use the On-line certificates to verify such signatures.

If the total received reply messages by the new joining node within a timeout period determined by the *ConfigTimer* are less than the threshold K then it repeats the *Autoconfiguration Service Requesting* using a higher HopLimit value. Otherwise, it starts the *Coalition selection* and *the On-line certificate requesting* procedures.

```
1:  const  n, RC_thresh  : integer
2:  int    RC            : array[0..n-1] {initialy RC=0}
3:  receive Config_Request from node P_i →
4:      if (P_i ∈ BL) then
5:          discard message;
6:      else if SigCheck(Config_Request) then
7:              if (RC[i] < RC_thresh) then
8:                  RC[i] := RC[i] + 1;
9:                  send Config_Replay to node P_i;
10:             else
11:                 discard message;
12:                 set node P_i as malicious and publish accusation;
13:             fi
14:         else
15:             discard message;
16:         fi
17:     fi
```

Figure 8. Algorithm 4 : Processing a Config_Request message by a Co-signer

### 4.4.3. Selection of Co-signers

The new joining node checks the validity of each reply message; it verifies that the received On-line CA's public key is the same one for at least *k* responders and saves it. Then, it starts selecting a subset (coalition) of co-signers and requesting an On-line certificate. The procedure is summarized in the following steps (the algorithm is given in Figure 9):

```
1:   bool ready = true;
2:   MaliciousServerTable = Ø;
3:   ready →
4:     select closest servers excluding malicious ones;
5:     select freeHostID;
6:     send Config_Cert_Request to selected servers;
7:     start ConfigCertTimer;
8:     ready := false;
9:   receive Config_Cert_Replay from combiner →
10:    if (SigCheck(Config_Cert_Replay)) then
11:      stop ConfigCertTimer;
12:      start registration;
13:    else
14:      discard message;
15:    fi
16:  receive Config_Alert →
17:    if (SigCheck(Config_Alert)) then
18:      update MaliciousServerTable;
19:      stop ConfigCertTimer;
20:      ready :=true;
21:    else
22:      discard message;
23:    fi
24:  timeout (ConfigCertTimer) →
25:    nodestate = unconfigured;
26:    return;
```

Figure 9. Algorithm 5 : Procedure of Co-signers coalition selection

- *Step1:* It selects among the closest responding nodes a coalition of at least K nodes according to the received HopLimit values appearing in the *Config_Reply* messages.

- *Step2:* It chooses randomly a lowest free Host-ID common to all the members of the selected coalition.

- *Step3:* It unicasts to these members a *Config_Cert_Request* message including the list of the coalition members, the chosen lowest free Host-ID, its 'Off-line Public Key Certificate' and its signature, expecting reception of a *Config_Cert_Reply* message from the combiner within a timeout period determined by the *ConfigCertTimer.*

### 4.4.4. Threshold Signature of the On-line Certificate

Each member in the coalition checks the validity of the *Config_Cert_Request* message, looks in its *CRL* and *BL* tables if no member of the coalition is malicious nor his public key is revoked. If this holds, then each member starts the threshold signature protocol providing an 'On-line Joint IP address and Public Key Certificate' for the requester. The member of the coalition with the lowest IP address will act as the combiner of the partial signatures, replies to the requester by a *Config_Cert_Reply* message, and has in addition the task to inform all nodes by a *Config_Advert* message that an IP address has been attributed to the node in question. Then, all nodes increment its *Requester Counter (RC)* and delete this address from the *FAT* and save it in the *PAT*. Hence, a new coming node will not have the possibility of choosing this address.

If a malicious node has been discovered among the coalition members, a *Config_Alert* message is sent to the honest members of the coalition and to the new joining node. This message includes the list of approved malicious members and/or the list of approved revoked public keys. The algorithm for this processing is shown in Figure 10.

The new joining node checks the correctness of this information by means of the On-line Certificate Authority's public key. Hence, it will be able to isolate the misbehaving nodes (either the node sending the *Config_Alert* message or nodes appearing in this message). Subsequently, the requester performs a new coalition selection while excluding the malicious nodes (see algorithm in Figure 9).

```
1:  receive Config_Cert_Request from P_i →
2:      if (SigCheck(Config_Cert_Request)) then
3:          if (requested IP Address ∈ RAT) then
4:              send Config_Error to P_i ;
5:          else if (there is a malicious node among the
                    selected servers ∈ BL or ∈ CRL)   then
6:              send Config_Alert to P_i and other servers;
7:          else
8:              call DPKI Manager; (certificate issuing)
9:          fi
10:     fi
11:     else
12:         discard message;
13:     fi
```

Figure 10. Algorithm 6 : Processing a certification request by a co-signer

### 4.4.5. MANET Registration

The new joining node, after receiving its 'On-line Joint IP address and Public Key Certificate', verifies its validity and proceeds to registration. This registration is indispensable for a node to be considered as member of the MANET. In doing so, it broadcasts a *Config_Register* message to all nodes in MANET using the site-local scope all-nodes multicast address FF05::1 as destination address. The *Config_Register* message must include the new node's 'On-line Joint IP address and Public Key Certificate' and the signature of the whole IPv6 packet. This affirms that the message has been sent to all nodes. This request must be processed by each node without any acknowledgement.

A conflict can occur by the acquisition of the same IP address, when two or more nodes try to join the network simultaneously and choose the same lowest Host-ID. Note that the registration of a same IP address can take place only during a limited interval of time. This conflict is resolved based on node's identities and the *RegistrationTimer*.

On one hand, the node which has broadcast a *Config_Register* message for a given IP address, and which receives, during the interval of time defined by the *RegistrationTimer*, a *Config_Register* message for the same IP address from another node must carry out the resolution procedure.

220

```
1:   bool ready = true;
2:   ready →
3:     broadcast Config_Register;
4:     start RegistrationTimer;
5:     ready := false;
6:   receive Config_Register from node P_i →
7:     if (SigCheck(Config_Register)) then
8:        if (my identity < P_i's identity) then
9:           send Config_Error to P_i;
10:       else
11:          nodestate := unconfigured;
12:          stop RegistrationTimer;
13:          return;
14:       fi
17:    else
18:       discard message;
19:    fi
20:  receive Config_Error from node P_i →
21:    if (SigCheck(Config_Error)) then
22:       if (my identity < P_i's identity) then
23:          send Config_Error to P_i;
24:       else
25:          nodestate := unconfigured;
26:          stop RegistrationTimer;
27:          return;
28:       fi
29:    else
30:       discard message;
31:    fi
32:  timeout (RegistrationTimer) →
33:    nodestate = configured;
34:    return;
```

Figure 11. Algorithm 7 : MANET Registration procedure and conflict resolution

In this procedure, the receiving node compares its identity with that of the sender. If it has the lowest identity, it maintains this IP address and unicasts a *Config_Error* message to the other node. Otherwise, it repeats the *Autoconfiguration Service Requesting* and chooses another IP address. The algorithm for this procedure is shown in Figure 11.

On the other hand, each node in the MANETs upon receiving *Config_Register* message, start the *RegistrationTimer* and delays the registration to its expiration. If another *Config_Register* message for the same IP Address is received before the timer elapses, the node with lowest identity will be registered. The algorithm for this processing is shown in Figure 12.

**4.5. Node Departure**

When a configured node leaves the network, it keeps its IP address. Next time, if it wishes to join the network, it can use this address as long as its On-line certificate is valid. If the certificate has expired, it must request a new address hence a new on-line certificate using its Off-line certificate.

**4.6. Network Partition and Merge**

MANETs may split in two or more partitions due to nodes mobility. To detect this partitioning, we adopt the protocol described in [6] in which periodic broadcast messages are performed by the node of lowest IP address. In our scheme, we define a partition as a subset of nodes with a size equal or greater than threshold K, and each partition has a unique identifier defined by a 4-tuple: the lowest IP address in use in the partition, the network prefix, the subnet (s) in use the

International Journal of Computer Networks & Communications (IJCNC) Vol.3, No.4, July 2011

partition and the network's public key. When a group of less than K nodes is dislocated from the MANET, it will not be considered as a partition. The arrival of other *unconfigured* nodes should help this group to create a new independent network according to protocol TCSAP if they remain isolated from the parent network.

```
1:   int LowestID = 0;
2:   receive Config_Register →
3:     if (Config_Register is a broadcast message) then
4:       if (SigCheck(Config_Register)) then
5:         if (the requester IP Address ∈ PAT) then
6:           if (there is a registration in progress for this
                 same IP Adress) then
7:             LowestID := lowest identity of the
                             requesters for the same IP
8:           Adress;
9:           else
10:            start RegistrationTimer;
11:            LowestID := requester's identity;
12:          fi
13:        else
14:          discard message;
15:        fi
16:      else
17:        discard message;
18:      fi
19:    else
20:      discard message;
21:    fi
22:  timeout (RegistrationTimer) →
23:    register the IP Adress in the RAT with LoewstID;
       delete the Ip Adress fron the PAT;
```

Figure 12. Algorithm 8 : Processing a registration request by each node in the network

Let us consider a network splitting in two or more partitions. The identifier of the partition which will contain the node with lowest IP address will stay unchanged. This partition will continue to provide the autoconfiguration service without any changes (i.e. no address clean up and no subnet change). Other partitions must generate new partition identifiers. To do so, the node with lowest IP address in use in the partition selects randomly K nodes to set up a new subnet. Thus, the new identifier will be represented by the following tuple: the new lowest IP address in use in the partition, the parent network prefix, the subnet (s) in use the partition (including the new one) and the parent network's public key. Each one of these partitions will provide the autoconfiguration service using only new subnet. There will not be any address clean up. When partitions' merging occurs, the node with lowest IP address in each partition collects a threshold signature of the partition state information and broadcast a message including this state information and this signature. Each node updates its state information according to this data resulting in global state information of the whole network. The case where two independent networks merge is not really an interesting problem since practically there will be no address conflict due to the very weak collision probability of networks prefix and networks public key.

222

## 5. SECURITY DISCUSSION

In this work, we have adopted a threshold cryptographic approach (N≥2K-1) to achieve for our scheme security and robustness in the presence of (K-1) faults. Unlike the previous approaches, which rely on a single node either in providing auto-configuration service, in our scheme the service is initiated and provided by at least k arbitrary honest nodes. Consequently, we avoid any single point of failure or trusted party. Moreover, our scheme is totally distributed over the whole network, and a new joining node does not need any particular distribution of its neighbors to be initiated with network and security parameters. Hence, the service availability is guaranteed ubiquitously. The mechanism of mutual authentication with Off-line certificates allows the servers to authenticate the requester, that is only legitimate nodes can take part in the network, but also the requester to authenticate the servers to prevent Man-In-the-Middle Attack. However, malicious nodes may be present among the servers selected by the requester. For this reason, Config_Alert messages are used to prevent malicious nodes from providing or disturbing the auto-configuration service. Threshold signature verification should also be used to isolate misbehaving nodes that are not yet in Black List. The requester may also be malicious, the Requester Counter (RC) and the registration mechanism can efficiently thwart Exhaustion Address Space and Sybil Attacks, the only possible ones in this case. The Traffic overload DoS Attack is prevented by the maximum authorized HopLimit (less than the threshold k) used in Config_Request messages. The mechanism of assigning an IP address by a coalition instead of a single entity solves the problem of Conflict Address Attack present in both stateful and stateless earlier approaches. The concept of 'On-line Joint IP address and Public Key Certificate' we introduced in our scheme represents, in the other hand, an effective mechanism to thwart IP Spoofing Attack. A malicious node which wants to spoof either an unused IP address or an already assigned IP address must hold an 'On-line Joint IP address and Public Key Certificate' in which its public key is bind to the spoofed IP address. Hence, instead of the limitations of the various solution approaches analyzed in [13], none of the attacks quoted in section 1 appear to break our proposal.

## 6. SIMULATION EXPERIMENTS

Simulation experiments were performed using the network simulator NS-2 [14] with CMU mobility extensions to evaluate the performance of our protocol in terms of configuration latency and communication overhead. The configuration latency metric represents the average delay for a new joining node to obtain an 'On-line Joint IP address and Public Key Certificate'. This includes all possible delays caused by the messages exchanges, timeouts and cryptographic primitives. The communication overhead metric represents the number of control packets transmitted during the auto-configuration process. The protocol TCSAP is implemented within NS-2 using C++, by creating a new agent TCSAP. The cryptographic primitives were simply modeled by delays. We used the results published by [15] for a 1.83 GHz Dual Core Intel processor, under Windows Vista (32 bits mode). We considered the RSA-2048 algorithm and the ECDSA-233 algorithm respectively for ordinary signature and threshold signature.

### 6.1. Simulation Scenarios and Parameters

The random waypoint mobility model was used. The simulation time was set to 120 seconds. We used the AODV routing protocol. Each data point represents an average value of five runs with the same settings, but different randomly generated topology and mobility scenarios.

The following sets of simulation were performed.

a) Impact of threshold K on Initialization Delay: We aim in this simulation to measure the delay of MANET initialization. We define this delay as the passing time between the moment of the first emitted *Init_Start* message and the moment where the MANET is completely initialized. Consequently, no motion was applied in this scenario. The coordinates of the nodes were selected in order to have each time the desired number of founding nodes which corresponds to the value of threshold K=3,4,5,8,10.

b) Varying network density: We study here the effect of the network density on Auto-configuration latency and communication overhead. The area of the network was set to 1000m*1000m, for the 15, 25, 50, 75 and 100 node populations, ensuring respectively 15, 25, 50, 75, and 100 (nodes/km2) for the network density. The simulations were performed for different values of threshold K. No motion was applied in this scenario.

c) Varying network mobility: we examine the protocol efficiency when the mobility of nodes increases. A network area of 1000m*1000m with 50 nodes is simulated for different values of threshold K. We vary the maximum node speed from 0 to 50 m/s; pause time is set to 0, according to the following command (example for 20 m/s node speed):

Setdest –v2 –n 50 –s 1 –m 20 –M 20 –t 120 –P 1 –p 0 –x 1000 –y 1000

### 6.2. Simulation Results

- Initialization Delay:

It was observed (Figure 13) that the mean Initialization Delay increases with the increase of threshold K. The TCSAP Initialization protocol and the joint verifiable random secret sharing protocol executed in MANET Initialization phase are based on a high number of messages exchanges and a high computational time for cryptographic primitives. These parameters are all the more large as the threshold value is larger. However, the measured Initialization Delay remains practical for low threshold values (less than 2 seconds for K=3,4,5).

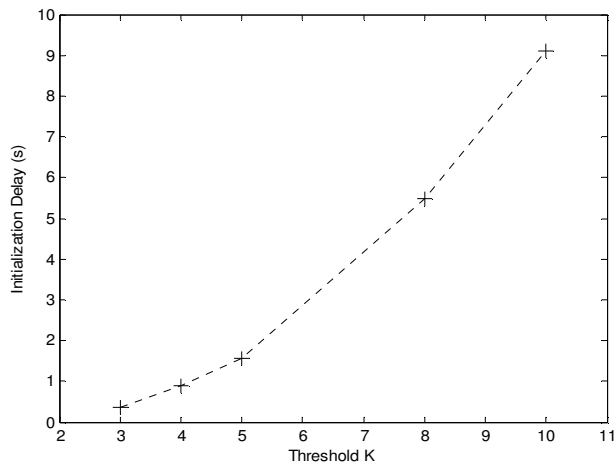

Figure 13 . Initialization Delay vs Threshold

- Auto-configuration Latency:

1. Impact of network density: Figure 14 shows an increase in Auto-configuration latency when the network density is low (below 25 nodes/km2), in particular for the high threshold values. The mean value of latency is less than one second. The minimum was observed at (25 nodes/km2). But again, from this point latency increases linearly with respect to density. Latency increases also with respect to the threshold parameter.

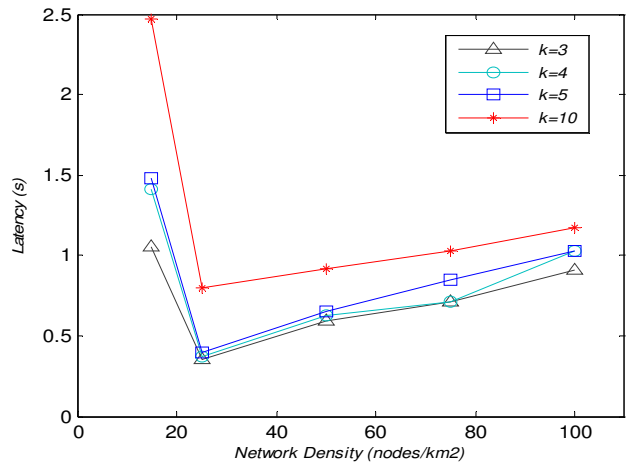

Figure 14. Configuration Latency vs Network Density

2. Impact of mobility: It was observed that node mobility has no significant effect on latency (Figure 15). This was because the simulated speeds were lower than 50m/s, and that the mean latency is less than 1 second, for such delay a node movement does not exceed 50m, and this in most time does not break links. In some particular cases, the mobility may have positive/ negative impact on latency.

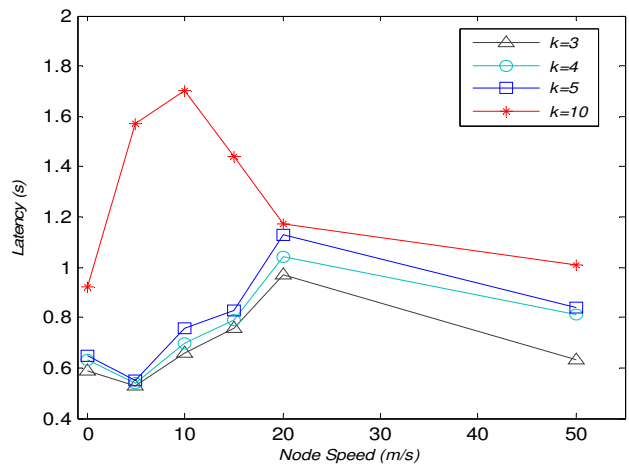

Figure 15. Configuration Latency vs Node Mobility

- Communication Overhead:

1. Impact of network density: In networks with high density, there are more nodes in the neighbourhood of the new joining node, and all reply to its autoconfiguration service requesting, leading to a higher number of messages exchange. For this raison, we observe in (Figure 16) an increasing in overhead when density increases. Note that this will increase also latency.

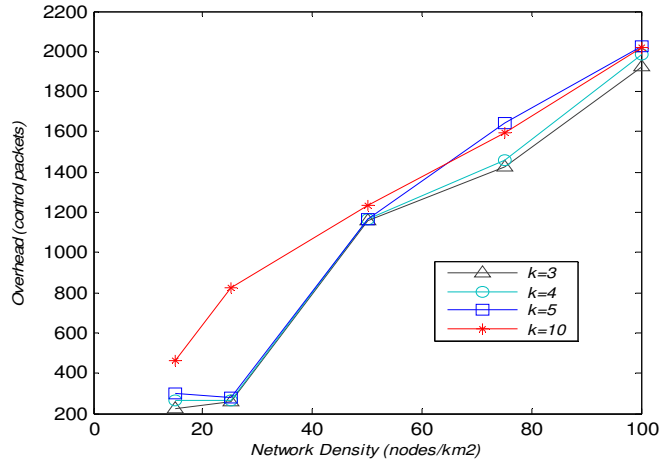

Figure 16. Communication Overhead vs Network Density

2. Impact of mobility: For the same reasons provided above, the node mobility has no significant effect on overhead. (Figure 17)

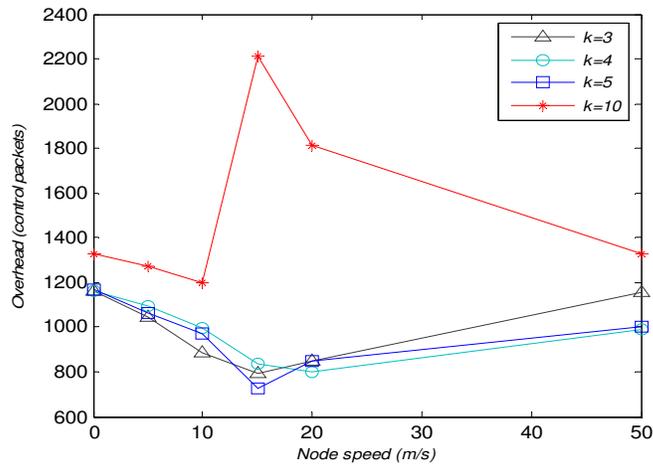

Figure 17. Communication Overhead vs Node Mobility

## 7. CONCLUSIONS

The TCSAP protocol proposed in this paper achieves IPv6 stateful dynamic configuration for MANETs. Our solution provides both security and robustness and overcomes all the limitations of the previously proposed approaches while still ensuring the timely IP address allocation. Furthermore, instead of others approaches which use separate mechanisms for network parameters configuration and security parameters configuration, our scheme achieves the same purpose at once, which make it efficient in terms of latency and communication overhead as shown by NS2 simulation results.

## REFERENCES

[1] F. Buiati, R. S. Puttini, and R. T. de Sousa Jr., "A Secure Autoconfiguration Protocol for MANET Nodes". *Lecture Notes in Computer Science*, v. 3158, pp. 108-121, 2004.

[2] Ana Cavalli, Jean-Marie Orset, "Secure hosts autoconfiguration in mobile ad hoc networks", *ICDCSW*.2004, pp. 809-814.

[3] Pan Wang, Douglas S. Reeves, Peng Ning, "Secure Address Autoconfiguration for Mobile Ad Hoc Networks", *MOBIQUITOUS* 2005. pp. 519-522.

[4] Shenglan Hu, Chris J. Mitchell, "Improving IP address autoconfiguration security in MANETs using trust modelling", *Mobile Ad-hoc and Sensor Networks - First International Conference*, MSN 2005, pp. 83-92

[5] André Langer and Tom Kühnert, "Security issues in Address Autoconfiguration Protocols: An improved version of the Optimized Dynamic Address Configuration Protocol". *archiv.tu-chemnitz.de*, 2007

[6] S. Nesargi and R. Prakash, "MANETconf: Configuration of Hosts in a Mobile Ad Hoc Network", *IEEE INFOCOM* 2002, June 2002

[7] J. Kong, P. Zerfos, H. Luo, S. Lu, and L. Zhang. Providing Robust and Ubiquitous Security Support for MANET. *In IEEE International Conference on Network Protocols*, pages 251.260, Nov. 2001.

[8] Di Crescenzo, G. Arce and R. Ge. Threshold Cryptography in Mobile Ad Hoc Networks. *In Springer Berlin / Heidelberg editor, Security in Communication Networks,* volume 3352 of Lecture Notes in Computer Science, Springer 2005.

[9] T. Pedersen, A threshold Cryptosystem without a Trusted Party, *in Proc. of Eurocrypt* 91.

[10] A. Shamir. How to Share a Secret. *Communications of the ACM*, 22(11):612–613, 1979

[11] M. Hwang and T. Chang. Threshold Signatures: Current Status and Key Issues. *International Journal of Network Security*, Vol.1, No.3, PP.123–137, Nov. 2005

[12] M. G. Gouda, "Elements of Network Protocol Design," *John Wiley and Sons*, 1998

[13] A. Abdelmalek, M. Feham and A. Taleb-Ahmed. On Recent Security Enhancements to Autoconfiguration Protocols for MANETs: Real Threats and Requirements. *International Journal of Computer Science and Network Security*, Vol.9, No.4, PP.401–407, April 2009

[14] The Network Simulator manual, *The NS2 homepage http://www.isi.edu/nsnam/ns*

[15] Speed Comparison of Popular Crypto Algorithms, http://www.cryptopp.com

[16] C. Crépeau and C.R. Davis. A Certificate Revocation Scheme for Wireless Ad Hoc Networks. *Proceedings of ACM Workshop on Security of Ad Hoc and Sensor Networks* (SASN '03), pp.54-61, 2003.